\documentclass[12pt]{article}
\usepackage{a4wide}
\usepackage{latexsym}
\usepackage{amsmath}
\usepackage{amsfonts}
\usepackage{amscd}
\usepackage{shuffle}
\usepackage{cite}
\usepackage{graphicx}
\usepackage{float}

\usepackage{pslatex}
\usepackage[latin1,utf8]{inputenc}
\usepackage[OT2,T1]{fontenc}

\usepackage{ifthen}

\newcommand{\bq}{\begin{eqnarray}}
\newcommand{\eq}{\end{eqnarray}}
\newcommand{\eps}{\varepsilon}

\begin{document}

\thispagestyle{empty}

\begin{flushright}
  MITP/16-077
% \\ version of \today
\end{flushright}

\vspace{1.5cm}

\begin{center}
  {\Large\bf Relations for Einstein-Yang-Mills amplitudes from the CHY representation\\
  }
  \vspace{1cm}
  {\large Leonardo de la Cruz, Alexander Kniss and Stefan Weinzierl\\
\vspace{2mm}
      {\small \em PRISMA Cluster of Excellence, Institut f{\"u}r Physik, }\\
      {\small \em Johannes Gutenberg-Universit{\"a}t Mainz,}\\
      {\small \em D - 55099 Mainz, Germany}\\
  } 
\end{center}

\vspace{2cm}

% abstract -------------------------------------------------------------------------
\begin{abstract}\noindent
  {
We show that a recently discovered relation, which expresses tree-level single trace Einstein-Yang-Mills amplitudes
with one graviton and $(n-1)$ gauge bosons as a linear combination of pure Yang-Mills tree amplitudes with
$n$ gauge bosons, can be derived from the CHY representation.
In addition we show that there is a generalisation, which 
expresses tree-level single trace Einstein-Yang-Mills amplitudes
with $r$ gravitons and $(n-r)$ gauge bosons as a linear combination of pure Yang-Mills tree amplitudes with
$n$ gauge bosons.
We present a general formula for this case.
   }
\end{abstract}

\vspace*{\fill}

% main text ------------------------------------------------------------------------
\newpage

% ----------------------------------------------------------------------------------
\section{Introduction}
\label{sect:intro}

On the one hand, relations between amplitudes in gauge theories are useful for reducing the number of independent 
amplitudes and have an application in the simplification of multi-leg amplitudes. 
On the other hand, relations between amplitudes in different theories are very interesting, as they connect two otherwise
unrelated theories.
A prominent example is the relation between graviton amplitudes in perturbative quantum gravity
and gauge theory amplitudes, provided by the KLT relations \cite{Kawai:1985xq},
BCJ duality \cite{Bern:2010ue} or the CHY representation \cite{Cachazo:2013iea}.

In this paper we are concerned with amplitudes in Einstein-Yang-Mills (EYM) theory.
These amplitudes consist of gauge bosons and gravitons.
Tree-level Einstein-Yang-Mills amplitudes were studied in \cite{Chen:2010ct} using BCFW recursion relations.
A CHY representation of Einstein-Yang-Mills amplitudes 
was worked out in refs.~\cite{Cachazo:2014nsa,Cachazo:2014xea}. 
In the context of string theory Einstein-Yang-Mills amplitudes have been studied extensively 
using the so called ``disk  relations'' \cite{Stieberger:2014cea,Stieberger:2015qja,Stieberger:2015kia,Stieberger:2015vya,Stieberger:2016lng,Chiodaroli:2014xia,Chiodaroli:2015rdg,Chiodaroli:2016jqw}.
In a recent paper, Stieberger and Taylor \cite{Stieberger:2016lng} presented a relation
relating tree-level single trace Einstein-Yang-Mills amplitudes
with one graviton and $(n-1)$ gauge bosons to a linear combination of pure Yang-Mills tree amplitudes with
$n$ gauge bosons, by considering the low-energy limit of string theory
and by taking appropriate collinear and soft limits.
They raised the question, if this relation can be derived from the CHY representation.
In this paper we show that this is indeed the case and deduct this relation from the CHY representation.
Our derivation directly leads to the generalisation towards
$r$ gravitons and $(n-r)$ gauge bosons.
We provide a general formula,
which relates tree-level single trace Einstein-Yang-Mills amplitudes
with $r$ gravitons and $(n-r)$ gauge bosons to a linear combination of pure Yang-Mills tree amplitudes with
$n$ gauge bosons.\footnote{During the preparation of this manuscript, the paper \cite{Nandan:2016pya} appeared on the arxiv
with similar content, giving an explicit generalisation with up to three gravitons.}

The CHY representation is based on the scattering equations \cite{Cachazo:2013gna,Cachazo:2013hca,Cachazo:2013iea}
and has triggered significant interest in the 
community \cite{Dolan:2013isa,Dolan:2014ega,He:2014wua,Naculich:2014naa,Naculich:2015zha,Naculich:2015coa,Weinzierl:2014ava,Kalousios:2013eca,Weinzierl:2014vwa,Lam:2014tga,Monteiro:2013rya,Cachazo:2015nwa,Baadsgaard:2015voa,Baadsgaard:2015hia,Baadsgaard:2015ifa,delaCruz:2015dpa,delaCruz:2015raa,Sogaard:2015dba,Bosma:2016ttj,Huang:2015yka,Cardona:2016gon,Gomez:2016bmv,Cardona:2016bpi,Cardona:2016wcr}.
In addition, there are interesting connections with string theory \cite{Bjerrum-Bohr:2014qwa,Mason:2013sva,Berkovits:2013xba,Gomez:2013wza,Adamo:2013tsa,Geyer:2014fka,Casali:2014hfa,Geyer:2015bja}
and gravity \cite{Schwab:2014xua,Afkhami-Jeddi:2014fia,Zlotnikov:2014sva,Kalousios:2014uva,White:2014qia,Monteiro:2014cda,delaCruz:2016wbr,White:2016jzc}.

This paper is organised as follows: 
In section~\ref{sect:notation} we introduce our conventions and state the relation
which we would like to show.
In section~\ref{sect:CHY} we review the CHY construction 
and give the building blocks for the Einstein-Yang-Mills amplitudes. 
In section~\ref{sect:derivation} we derive the 
the Einstein-Yang-Mills relations from the CHY representation. 
We provide a generalisation for more than one graviton in section~\ref{sect:generalisations}.
Finally, our conclusions are given in section~\ref{sect:conclusions}.

% ----------------------------------------------------------------------------------
\section{Notation}
\label{sect:notation}

We denote gauge bosons by $g$ and gravitons by $h$.
The two polarisation states of a gauge boson are denoted $g^+$ and $g^-$, the two polarisation states
of the graviton by $h^{++}$ and $h^{--}$.
Our convention for the normalisation of the generators of the gauge group is
\bq
 \mathrm{Tr}\; T^a T^b & = & \frac{1}{2} \delta^{ab}.
\eq
In this paper we are on the one hand concerned with tree-level amplitudes in pure Yang-Mills theory.
These amplitudes can be decomposed into sums of
group-theoretical factors multiplied by kinematic functions called primitive amplitudes. 
For the $n$-gauge boson tree amplitude this decomposition reads
\bq
{\cal A}_{n}^{\mathrm{YM}}(g_1,g_2,...,g_n) 
 & = & g^{n-2} \sum\limits_{\sigma \in S_{n}/Z_{n}} 
 2 \; \mbox{Tr} \left( T^{a_{\sigma(1)}} ... T^{a_{\sigma(n)}} \right)
 \;\;
 A_{n}^{\mathrm{YM}}\left( \sigma, p, \eps \right),
\eq
where the sum is over all non-cyclic permutations $\sigma$ of the external legs.
The prefactor $g$ denotes the coupling of Yang-Mills theory.
The primitive amplitudes $A_{n}^{\mathrm{YM}}$ are gauge invariant 
and each primitive amplitude has a fixed cyclic order of the external legs,
specified by $\sigma=(\sigma_1,...,\sigma_n)$.
The $n$-tuple of external momenta corresponding to $(g_1,g_2,...,g_n)$ is denoted
by $p=(p_1,...,p_n)$ and
the $n$-tuple of external polarisations by $\eps=(\eps_1^{\lambda_1},...,\eps_n^{\lambda_n})$
with $\lambda_j \in \{+,-\}$.
It is sometimes convenient to use an alternative notation for the arguments of the primitive amplitudes:
\bq
 A_{n}^{\mathrm{YM}}\left( p_{\sigma_1}^{\lambda_{\sigma_1}}, ..., p_{\sigma_n}^{\lambda_{\sigma_n}}\right)
 & = &
 A_{n}^{\mathrm{YM}}\left( \sigma, p, \eps \right)
\eq
We will use both notations interchangeable.

On the other hand we are concerned with tree-level amplitudes in Einstein-Yang-Mills theory
with $(n-1)$ gauge bosons and one graviton.
These may be written as
\bq
\lefteqn{
{\cal A}_{n,n-1}^{\mathrm{EYM}}(g_1,...,g_{n-1},h_n) 
 =
} & & \nonumber \\
 & &
 \frac{\kappa}{4} g^{n-3} \sum\limits_{\sigma \in S_{n-1}/Z_{n-1}} 
 2 \; \mbox{Tr} \left( T^{a_{\sigma(1)}} ... T^{a_{\sigma(n-1)}} \right)
 \;\;
 A_{n,n-1}^{\mathrm{EYM}}\left( p_{\sigma_1}^{\lambda_{\sigma_1}}, ..., p_{\sigma_{n-1}}^{\lambda_{\sigma_{n-1}}}, p_n^{\lambda_n \lambda_n} \right)
 + ...,
\eq
where the dots stand for contributions with two or more traces.
For amplitudes in Einstein-Yang-Mills theory we use the notation ${\cal A}_{n,r}^{\mathrm{EYM}}$, indicating that the amplitude
contains $r$ external gauge bosons and $(n-r)$ external gravitons.
The coupling of the graviton is $\kappa/4$, where $\kappa=\sqrt{32\pi G_N}$ in the Gau{\ss} unit system
and $\kappa=\sqrt{8 G_N}$ in natural units.
In this paper we are interested in the tree-level single trace amplitudes $A_{n,n-1}^{\mathrm{EYM}}$.
In principle we may compute these amplitudes from the Lagrange density
\bq
 {\mathcal L}_{\mathrm{EYM}}
 & = &
 - \frac{2}{\kappa^2} \sqrt{-\det(g_{\mu\nu})} R
 - \frac{1}{4} \sqrt{-\det(g_{\mu\nu})} g^{\mu_1\mu_2} g^{\nu_1\nu_2} F_{\mu_1\nu_1}^a F^a_{\mu_2\nu_2},
\eq
by perturbatively expanding around the flat Minkowski metric $\eta_{\mu\nu}$
\bq
 g_{\mu\nu}
 & = & 
 \eta_{\mu\nu} + \kappa h_{\mu\nu}.
\eq
Alternatively we may compute the amplitudes $A_{n,n-1}^{\mathrm{EYM}}$ with the help of the CHY representation, which we will
review in section~\ref{sect:CHY}.
The polarisation of the graviton is described by a product of two polarisation vectors for gauge bosons:
\bq
\label{graviton_polarisation}
 \eps_n^{\pm \pm} & = & \eps_n^\pm \eps_n^\pm
\eq
Also in the case of Einstein-Yang-Mills amplitudes it will be convenient to introduce a second notation for the
arguments of the amplitudes. We write
\bq
 A_{n,n-1}^{\mathrm{EYM}}\left( p_{\sigma_1}^{\lambda_{\sigma_1}}, ..., p_{\sigma_{n-1}}^{\lambda_{\sigma_{n-1}}}, p_n^{\lambda_n \lambda_n} \right)
 & = &
 A_{n,n-1}^{\mathrm{EYM}}\left( \sigma, p, \eps, \tilde{\eps} \right),
\eq
where $\sigma \in S_{n-1}$ is a permutation of $(n-1)$ elements, $p=(p_1,...,p_n)$ and $\eps=(\eps_1^{\lambda_1},...,\eps_n^{\lambda_n})$ as before,
while $\tilde{\eps}=(\eps_n^{\lambda_n})$ contains only a single element.
In this way the two polarisation vectors for the graviton, appearing on the right-hand side of eq.~(\ref{graviton_polarisation}), 
are split between $\eps$ and $\tilde{\eps}$.

Recently, Stieberger and Taylor \cite{Stieberger:2016lng} obtained the relation
\bq
\label{EYM_relation}
 A_{n,n-1}^{\mathrm{EYM}}\left( p_{1}^{\lambda_1}, ..., p_{n-1}^{\lambda_{n-1}}, p_n^{\lambda_n \lambda_n} \right)
 = 
 -
 \sum\limits_{j=1}^{n-2}
 \left( 2 q_j \cdot \eps_n^{\lambda_n} \right)
 A_{n}^{\mathrm{YM}}\left( p_{1}^{\lambda_1}, ..., p_{j}^{\lambda_{j}}, p_n^{\lambda_n}, p_{j+1}^{\lambda_{j+1}}, ..., p_{n-1}^{\lambda_{n-1}} \right)
\eq
where
\bq
 q_j & = & \sum\limits_{k=1}^j p_k.
\eq
The minus sign on the right-hand side of eq.~(\ref{EYM_relation}) is related to the sign convention of the gauge coupling
within the field strength tensor.
Our convention is $F_{\mu\nu} = \partial_\mu A^a_\nu - \partial_\nu A^a_\mu + g f^{abc} A^b_\mu A^c_\nu$.
Eq.~(\ref{EYM_relation}) expresses the 
tree-level single trace Einstein-Yang-Mills amplitude
with one graviton and $(n-1)$ gauge bosons as a linear combination of pure Yang-Mills amplitudes with
$n$ gauge bosons.
In ref.~\cite{Stieberger:2016lng} the relation~(\ref{EYM_relation}) was derived from the low-energy limit of string theory
and by considering appropriate collinear and soft limits.
Ref.~\cite{Stieberger:2016lng} raised the question, if eq.~(\ref{EYM_relation}) can be derived within the framework
of the scattering equations.
In this paper we give an affirmative answer to this question and we derive eq.~(\ref{EYM_relation}) from the CHY representation.

% ----------------------------------------------------------------------------------
\section{The CHY representation}
\label{sect:CHY}

For massless particles the scattering equations read
\bq
 f_i\left(z,p\right) & = & 
 \sum\limits_{j=1, j \neq i}^n \frac{ 2 p_i \cdot p_j}{z_i - z_j}
 \;\; = \;\; 0.
\eq
For the variables $z_j$ we have $z_j \in {\mathbb C} \cup \{\infty\}$.
There are $(n-3)!$ inequivalent solutions for the $n$-tuples $z=(z_1,z_2,...,z_n)$.
Two solutions are called inequivalent if they are not related by a $\mathrm{PSL}(2,{\mathbb C})$ transformation.
The scattering equation formalism allows us to represent tree-level amplitudes in Yang-Mills theory and Einstein-Yang-Mills theory
as a sum over the inequivalent solutions of the scattering equations.
This is the CHY representation.

We define three ingredients: A Jacobian factor $J(z,p)$, a cyclic factor (or Parke-Taylor factor) $C_r(\sigma,z)$ 
and a polarisation factor $E_r(z,p,\eps)$.
Note that the Jacobian factor neither depends on the cyclic order $\sigma$ nor on the polarisations $\eps$.
The cyclic factor depends only on the cyclic order $\sigma$, but not on the polarisations $\eps$.
On the other hand, the polarisation factor depends on the polarisations $\eps$, but not on the cyclic order $\sigma$.

We start with the definition of the Jacobian factor $J(z,p)$.
It will be convenient to set $z_{ij} = z_i - z_j$.
We define a $n \times n$-matrix $\Phi(z,p)$ with entries
\bq
 \Phi_{ab}\left(z,p\right)
 & = &
 \frac{\partial f_a\left(z,p\right)}{\partial z_b}
 \;\; = \;\;
 \left\{
 \begin{array}{cc}
 \frac{2 p_a \cdot p_b}{z_{ab}^2} & a \neq b, \\
 - \sum\limits_{j=1, j \neq a}^n \frac{2 p_a \cdot p_j}{z_{aj}^2} & a=b. \\
 \end{array}
 \right.
\eq
Let $\Phi^{ijk}_{rst}(z,p)$ denote the $(n-3)\times(n-3)$-matrix,
where the rows $\{i,j,k\}$ and the columns $\{r,s,t\}$ have been deleted.
We set
\bq
 \det{}' \; \Phi\left(z,p\right)
 & = &
 \left(-1\right)^{i+j+k+r+s+t}
 \frac{\left|\Phi^{ijk}_{rst}(z,p)\right|}{\left(z_{ij}z_{jk}z_{ki}\right)\left(z_{rs}z_{st}z_{tr}\right)}.
\eq
With the above sign included,
the quantity $\det{}' \; \Phi(z,p)$ is independent of the choice of $\{i,j,k\}$ and $\{r,s,t\}$.
The Jacobian factor is given by
\bq
\label{def_jacobian}
 J\left(z,p\right) & = &
 \frac{1}{\det{}' \; \Phi\left(z,p\right)}.
\eq
Let us now consider a permutation $\sigma \in S_r$ with $2 \le r \le n$.
The cyclic factor (or Parke-Taylor factor) is given by
\bq
\label{def_cyclic_factor}
 C_r\left(\sigma,z\right)
 & = & 
 \frac{1}{z_{\sigma_1 \sigma_2} z_{\sigma_2 \sigma_3} ... z_{\sigma_r \sigma_1}}.
\eq
For $r=0$ we set $C_0(\sigma,z)=1$, for $r=1$ we set $C_1(\sigma,z)=0$.

Let us now turn to the polarisation factor $E_r(z,p,\eps)$. 
For a subset $\{i_1,i_2,...,i_r\} \subseteq \{1,2,...,n\}$ with $1 \le r \le n$ 
we denote the corresponding subset of polarisation vectors by $\eps'=(\eps_{i_1}^{\lambda_{i_1}}, ..., \eps_{i_r}^{\lambda_{i_r}})$.
We define a $(2r)\times(2r)$ antisymmetric matrix $\Psi(z,p,\eps')$ through 
\bq
\label{def_Psi_1}
 \Psi\left(z,p,\eps'\right)
 & = &
 \left( \begin{array}{cc}
 A & - C^T \\
 C & B \\
 \end{array} \right)
\eq
with
\bq
 A_{ab}
 = 
 \left\{ \begin{array}{cc}
 \frac{2 p_{i_a} \cdot p_{i_b}}{z_{i_a}-z_{i_b}} & a \neq b, \\
 0 & a = b, \\
 \end{array} \right.
 & &
 B_{ab}
 = 
 \left\{ \begin{array}{cc}
 \frac{2 \eps_{i_a} \cdot \eps_{i_b}}{z_{i_a}-z_{i_b}} & a \neq b, \\
 0 & a = b, \\
 \end{array} \right.
\eq
and
\bq
\label{def_Psi_3}
 C_{ab}
 & = &
 \left\{ \begin{array}{cc}
 \frac{2 \eps_{i_a} \cdot p_{i_b}}{z_{i_a}-z_{i_b}} & a \neq b, \\
 - \sum\limits_{j=1, j \neq i_a}^n \frac{2 \eps_{i_a} \cdot p_j}{z_{i_a}-z_j}  & a = b. \\
 \end{array} \right.
\eq
Note that the sum in the diagonal entries of $C_{ab}$ is over $(n-1)$ terms and not just $(r-1)$ terms.
For $1 \le r \le (n-2)$ we set 
\bq
 E_r\left(z,p, \eps' \right)
 & = &
 \mathrm{Pf} \; \Psi\left(z,p,\eps'\right).
\eq
For $r=0$ we set $E_0(z,p,\eps')=1$, for $r=n-1$ we set $E_{n-1}(z,p,\eps')=0$.
For the case $r=n$ the polarisation factor $E_n(z,p,\eps)$ is defined in terms of a reduced Pfaffian.
We denote for $1 \le i < j \le n$
by $\Psi^{ij}_{ij}(z,p,\eps)$ the $(2n-2)\times(2n-2)$-matrix 
where the rows and columns $i$ and $j$ of $\Psi(z,p,\eps)$ have been deleted.
We then set
\bq
 E_n\left(z,p,\eps\right)
 & = &
 \frac{\left(-1\right)^{i+j}}{2 z_{ij}} \mathrm{Pf} \; \Psi^{ij}_{ij}\left(z,p,\eps\right).
\eq
With all ingredients at hand we may now present the CHY representation of tree amplitudes
in Einstein-Yang-Mills theory.
We consider scattering amplitudes with $r$ gauge bosons and $(n-r)$ gravitons.
Without loss of generality we assume that the first $r$ particles are gauge bosons, while the
particles labelled from $(r+1)$ to $n$ are gravitons.
We denote by $\sigma$ a permutation of the gauge bosons (i.e. a permutation of $(1,...,r)$)
and we set
\bq
 \eps \;\; = \;\; \left( \eps_1^{\lambda_1}, ..., \eps_r^{\lambda_r}, \eps_{r+1}^{\lambda_{r+1}}, ..., \eps_n^{\lambda_n}  \right),
 & &
 \tilde{\eps} \;\; = \;\; \left( \eps_{r+1}^{\lambda_{r+1}}, ..., \eps_n^{\lambda_n} \right).
\eq
Then
\bq
\label{CHY_representation_n_r}
 A_{n,r}^{\mathrm{EYM}}\left(\sigma, p, \eps, \tilde{\eps} \right)
 & = &
 i
 \sum\limits_{\mathrm{solutions} \; j} 
 J\left(z^{(j)},p\right) \; C_r\left(\sigma, z^{(j)}\right) \; E_n\left(z^{(j)},p,\eps\right) \; E_{n-r}\left(z^{(j)},p,\tilde{\eps}\right),
\eq
where the sum is over the inequivalent solutions of the scattering equations.
A few special cases are worth mentioning explicitly: For $r=n$ we obtain the pure gauge boson amplitudes
\bq
\label{CHY_representation_YM}
 A_n^{\mathrm{YM}}\left(\sigma, p, \eps \right)
 & = &
 i
 \sum\limits_{\mathrm{solutions} \; j} 
 J\left(z^{(j)},p\right) \; C_n\left(\sigma, z^{(j)}\right) \; E_n\left(z^{(j)},p,\eps\right),
\eq
for $r=0$ the pure graviton amplitudes
\bq
 M_n^{\mathrm{E}}\left(p, \eps, \tilde{\eps} \right)
 & = &
 i
 \sum\limits_{\mathrm{solutions} \; j} 
 J\left(z^{(j)},p\right) E_n\left(z^{(j)},p,\eps\right) \; E_n\left(z^{(j)},p,\tilde{\eps}\right).
\eq
In this paper we are interested in Einstein-Yang-Mills amplitudes with $(n-1)$ gauge bosons and one graviton. 
These correspond to the case $r=n-1$.
Explicitly we have
\bq
 A_{n,n-1}^{\mathrm{EYM}}\left(\sigma, p, \eps, \tilde{\eps} \right)
 & = &
 i
 \sum\limits_{\mathrm{solutions} \; j} 
 J\left(z^{(j)},p\right) \; C_{n-1}\left(\sigma, z^{(j)}\right) \; E_n\left(z^{(j)},p,\eps\right) \; E_{1}\left(z^{(j)},p,\tilde{\eps}\right).
\eq
Note that eq.~(\ref{CHY_representation_n_r}) may also be written as
a multi-dimensional contour integral:
\bq
 A_n^{(0)}\left(\sigma,p,\eps\right)
 & = &
 i \oint\limits_{\mathcal C} d\Omega_{\mathrm{CHY}} \; 
 C_r\left(\sigma, z\right) \; E_n\left(z,p,\eps\right) \; E_{n-r}\left(z,p,\tilde{\eps}\right)
\eq
The measure $d\Omega_{\mathrm{CHY}}$ is defined by
\bq
 d\Omega_{\mathrm{CHY}}
 & = &
 \frac{1}{\left(2\pi i\right)^{n-3}}
 \frac{d^nz}{d\omega}
 \;
 \prod{}' \frac{1}{f_a\left(z,p\right)},
\eq
with
\bq
 \prod{}' \frac{1}{f_a\left(z,p\right)}
 & = & 
 \left(-1\right)^{i+j+k}
 z_{ij} z_{jk} z_{ki}
 \prod\limits_{a \neq i,j,k} \frac{1}{f_a\left(z,p\right)},
\eq
and
\bq
 d\omega
 & = &
 \left(-1\right)^{p+q+r}
 \frac{dz_p dz_q dz_r}{z_{pq} z_{qr} z_{rp}}.
\eq

% ----------------------------------------------------------------------------------
\section{Derivation of the relation from the CHY representation}
\label{sect:derivation}

In this section we derive the relation eq.~(\ref{EYM_relation}), relating
tree-level single trace Einstein-Yang-Mills amplitudes with one graviton
and $(n-1)$ gauge bosons to a linear combination of pure Yang-Mills amplitudes with $n$ gauge bosons from the
CHY representation.
It will be convenient to introduce a short-hand notation for the cyclic factors. We write
\bq
 C_r\left(1,2,...,r\right)
 & = &
 C_r\left( \left(1,2,...,r\right), z \right).
\eq
Our derivation is based on the following identity, reminiscent of eikonal factors:
\bq
\label{eq_eikonal}
 \sum\limits_{l=k}^{n-2} \frac{z_{l}-z_{l+1}}{\left(z_l-z_n\right)\left(z_n-z_{l+1}\right)}
 & = &
 \frac{z_k-z_{n-1}}{\left(z_k-z_n\right)\left(z_n-z_{n-1}\right)}
\eq
Eq.~(\ref{eq_eikonal}) is easily proven by repeated use of the identity
\bq
 \frac{z_k-z_{l+1}}{\left(z_k-z_n\right)\left(z_n-z_{l+1}\right)}
 +
 \frac{z_{l+1}-z_{l+2}}{\left(z_{l+1}-z_n\right)\left(z_n-z_{l+2}\right)}
 & = &
 \frac{z_k-z_{l+2}}{\left(z_k-z_n\right)\left(z_n-z_{l+2}\right)}.
\eq
In order to derive eq.~(\ref{EYM_relation}), we write down the left-hand side and the right-hand side of this relation
and insert the CHY representation for all amplitudes.
Let us start with the left-hand-side, suppressing all arguments which are not relevant to the discussion:
\bq
 \mathrm{lhs}
 & = &
 A_{n,n-1}^{\mathrm{EYM}}
 \;\; = \;\;
 i 
 \sum\limits_{\mathrm{solutions} \; j} 
 J \; C_{n-1}\left(1,...,n-1\right) \; E_n \; E_1
\eq
The polarisation factor $E_1$ is given by
\bq
 E_1 & = & - \sum\limits_{k=1}^{n-1} \frac{2 p_k \cdot \eps_n}{z_k-z_n}
 \;\; = \;\;
 - \sum\limits_{k=1}^{n-2} 2 p_k \cdot \eps_n \frac{z_k-z_{n-1}}{\left(z_k-z_n\right)\left(z_n-z_{n-1}\right)}.
\eq
For the second equality we first separated the term $k=n-1$ from the remaining sum, used momentum conservation
$p_{n-1}=-p_1-...-p_{n-2}-p_n$ within this term and the fact that $2 p_n \cdot \eps_n=0$.
We thus have
\bq
\label{eq_lhs}
 \mathrm{lhs}
 & = &
 - i 
 \sum\limits_{\mathrm{solutions} \; j} 
 J \; C_{n-1}\left(1,...,n-1\right) \; E_n 
 \sum\limits_{k=1}^{n-2} 2 p_k \cdot \eps_n \frac{z_k-z_{n-1}}{\left(z_k-z_n\right)\left(z_n-z_{n-1}\right)}.
\eq
Let us now turn to the right-hand side of eq.~(\ref{EYM_relation}).
\bq
 \mathrm{rhs}
 & = &
 -
 \sum\limits_{l=1}^{n-2}
 2 q_l \cdot \eps_n 
 \;
 A_{n}^{\mathrm{YM}}\left( p_{1}, ..., p_{l}, p_n, p_{l+1}, ..., p_{n-1} \right)
 \nonumber \\
 & = &
 -
 i
 \sum\limits_{\mathrm{solutions} \; j} 
 J \; E_n
 \sum\limits_{l=1}^{n-2}
 \sum\limits_{k=1}^l
 2 p_k \cdot \eps_n 
 \;
 C_n\left(1, ...,l,n,l+1,...,n-1\right),
\eq
where we inserted the CHY representation and used the definition of $q_l=\sum\limits_{k=1}^l p_k$.
We now exchange the order of the $l$- and $k$-summation and use
\bq
 C_n\left(1, ...,l,n,l+1,...,n-1\right)
 & = &
 \frac{z_l-z_{l+1}}{\left(z_l-z_n\right)\left(z_n-z_{l+1}\right)}
  C_{n-1}\left(1,...,n-1\right).
\eq
We obtain
\bq
\label{eq_rhs}
 \mathrm{rhs}
 & = &
 -
 i
 \sum\limits_{\mathrm{solutions} \; j} 
 J \; C_{n-1}\left(1,...,n-1\right) \; E_n
 \sum\limits_{k=1}^{n-2}
 2 p_k \cdot \eps_n 
 \sum\limits_{l=k}^{n-2}
 \frac{z_l-z_{l+1}}{\left(z_l-z_n\right)\left(z_n-z_{l+1}\right)}.
\eq
We now see that eq.~(\ref{eq_rhs}) equals eq.~(\ref{eq_lhs}) if for all $k\le (n-2)$ and $n \ge 3$ we have
\bq
 \sum\limits_{l=k}^{n-2}
 \frac{z_l-z_{l+1}}{\left(z_l-z_n\right)\left(z_n-z_{l+1}\right)}
 & = &
 \frac{z_k-z_{n-1}}{\left(z_k-z_n\right)\left(z_n-z_{n-1}\right)}.
\eq
This is precisely eq.~(\ref{eq_eikonal}) and completes the proof of eq.~(\ref{EYM_relation}).

% ----------------------------------------------------------------------------------
\section{Generalisations}
\label{sect:generalisations}

Let us now discuss generalisations of eq.~(\ref{EYM_relation}) towards tree-level single trace amplitudes
with $r$ gauge bosons and $(n-r)$ gravitons.
The CHY representation of these amplitudes, given in eq.~(\ref{CHY_representation_n_r}), provides us with guidance.
We first note that not all primitive Yang-Mills amplitudes $A_n^{\mathrm{YM}}(w,p,\eps)$ (where $w$ denotes the cyclic order) are independent.
Cyclic invariance, 
Kleiss-Kuijf relations \cite{Kleiss:1988ne}
and Bern-Carrasco-Johansson relations \cite{Bern:2008qj}
reduce the number of independent primitive Yang-Mills amplitudes to $(n-3)!$ independent ones.
We may fix three external particles at specified positions, which we choose without loss of generality
as $1$, $(n-1)$ and $n$.
We denote by $B$ the corresponding set of independent cyclic orders.
By inspection of eq.~(\ref{CHY_representation_n_r}) and eq.~(\ref{CHY_representation_YM})
we see that we may express the Einstein-Yang-Mills amplitudes with $(n-r)$ gravitons as a linear combination
of Yang-Mills amplitudes with cyclic order $w\in B$, provided we find coefficients $\alpha_w(\sigma,p,\tilde{\eps})$
such that
\bq
\label{linear_system}
 J\left(z^{(j)},p\right) C_r\left(\sigma, z^{(j)}\right) \; E_{n-r}\left(z^{(j)},p,\tilde{\eps}\right)
 & = &
 \alpha_w\left(\sigma,p,\tilde{\eps}\right) \; J\left(z^{(j)},p\right) C_n\left(w, z^{(j)}\right).
\eq
Note that we only need to find the coefficients $\alpha_w$ when the variables $z^{(j)}$ are a solution
of the scattering equations.
The linear system in eq.~(\ref{linear_system}) has a solution.
With the notation of \cite{delaCruz:2015raa} we define a $(n-3)! \times (n-3)!$-matrix $M^{\mathrm{red}}_{w j}$ through
\bq
\label{entries_M_w_j}
 M^{\mathrm{red}}_{w j}
 & = &
 J\left(z^{(j)},p\right) \; C_n\left(w, z^{(j)}\right),
\eq
and a vector $G_j$ through
\bq
 G_j & = &  J\left(z^{(j)},p\right) C_r\left(\sigma, z^{(j)}\right) \; E_{n-r}\left(z^{(j)},p,\tilde{\eps}\right).
\eq
Thus we seek a solution of the linear system of equations
\bq
 \alpha_w M^{\mathrm{red}}_{w j} & = & G_j.
\eq
The matrix $M^{\mathrm{red}}_{w j}$ has an inverse $N^{\mathrm{red}}_{j w}$, given
by
\bq
\label{def_N_red}
 N^{\mathrm{red}}_{j w}
 & = &
 \sum\limits_{v \in B}
 S\left[w | \bar{v} \right] C_n\left( \bar{v}, z^{(j)}\right),
\eq
where for $v=(l_1, l_2, ..., l_{n-2}, l_{n-1}, l_n)$ the cyclic order $\bar{v}$ is defined by
$\bar{v} = ( l_1, l_2, ..., l_{n-2}, l_{n}, l_{n-1} )$.
The momentum kernel $S[w_1|\bar{w}_2]$ is defined 
for $w_1=(l_1 ... l_n) \in B$ and $w_2=(k_1 ... k_n) \in B$
by \cite{Cachazo:2013gna,Kawai:1985xq,BjerrumBohr:2010ta,BjerrumBohr:2010hn}
\bq
 S\left[w_1|\bar{w}_2\right]
 & = &
 \left(-1\right)^n
 \prod\limits_{i=2}^{n-2} 
 \left[
        2 p_{l_1} \cdot p_{l_i} 
        + \sum\limits_{j=2}^{i-1} \theta_{\bar{w}_2}\left(l_j,l_i\right) 
                                  2 p_{l_j} \cdot p_{l_i} 
 \right],
\eq
with
\bq
\theta_{\bar{w}_2}\left(l_j,l_i\right)
 & = & 
 \left\{
 \begin{array}{ll}
  1 & \mbox{if $l_j$  comes before $l_i$ in the sequence $k_2,k_3,...,k_{n-2}$}, \\
  0 & \mbox{otherwise}. \\
 \end{array}
 \right.
 \nonumber
\eq
Thus the coefficients $\alpha_w(\sigma,p,\tilde{\eps})$ are given by
\bq
\label{def1_coeff_alpha}
 \alpha_w\left(\sigma,p,\tilde{\eps}\right)
 & = &
 G_j N^{\mathrm{red}}_{j w}.
\eq
We then obtain
\bq
\label{final_result}
 A_{n,r}^{\mathrm{EYM}}\left(\sigma, p, \eps, \tilde{\eps} \right)
 & = &
 \sum\limits_{w \in B}
 \alpha_w\left(\sigma,p,\tilde{\eps}\right) A_n^{\mathrm{YM}}\left(w,p,\eps\right),
\eq
which provides the generalisation of eq.~(\ref{EYM_relation})
towards an arbitrary number of gravitons.
Eq.~(\ref{final_result}) expresses a tree-level single trace Einstein-Yang-Mills amplitude with
$r$ gauge bosons and $(n-r)$ gravitons as a linear combination
of pure Yang-Mills tree amplitudes with $n$ gauge bosons.
Note that the expression on the right-hand side of eq.~(\ref{final_result}) is in a BCJ basis, while
the expression on the right-hand side of eq.~(\ref{EYM_relation}) is not in a BCJ basis and can be reduced with the
help of the BCJ relations.

Note that the coefficients $\alpha_w(\sigma,p,\tilde{\eps})$ may be written as
\bq
\label{def2_coeff_alpha}
 \alpha_w\left(\sigma,p,\tilde{\eps}\right)
 & = &
 \sum\limits_{v \in B}
 S\left[w | \bar{v} \right] 
 \oint\limits_{\mathcal C} d\Omega_{\mathrm{CHY}} \; 
 C_r\left(\sigma, z\right) \; 
 E_{n-r}\left(z,p,\tilde{\eps}\right) \; 
 C_n\left( \bar{v}, z\right).
\eq
The integral in eq.~(\ref{def2_coeff_alpha}) is a global residue.
The expressions for the coefficients $\alpha_w$ in eq.~(\ref{def1_coeff_alpha})
and eq.~(\ref{def2_coeff_alpha}) offer a compact representation valid for all $n$ and $r$,
at the expense of a sum over the solutions of the scattering equations.
Let us stress that due to the methods of \cite{Sogaard:2015dba,Bosma:2016ttj}, 
this sum can always be performed
without knowing the solutions of the scattering equations, giving a (in general lengthy)
rational function in the scalar products $2p_i p_j$, $2 p_i \eps_j$ and $2 \eps_i \eps_j$.
For two gravitons (corresponding to $r=n-2$) and three gravitons ($r=n-3$)
one recovers the explicit formulae of \cite{Nandan:2016pya}, once a common BCJ basis has been chosen.

Let us give an example: We consider the amplitude with $2$ gravitons and $2$ gauge bosons, thus
$n=4$ and $r=2$. The basis $B$ consists of a single element, which we may take as $w=(1,2,3,4)$. 
For the permutation of the two gauge bosons we take $\sigma=(1,2)$.
We obtain
\bq
 A_{4,2}^{\mathrm{EYM}}\left(\sigma, p, \eps, \tilde{\eps} \right)
 & = &
 t
 \left[ 
        2 \tilde{\eps}_3 \cdot \tilde{\eps}_4 
        + \frac{\left(2 \tilde{\eps}_3 \cdot p_2 \right) \left(2 \tilde{\eps}_4 \cdot p_1 \right)}{t}  
        + \frac{\left(2 \tilde{\eps}_3 \cdot p_1 \right) \left(2 \tilde{\eps}_4 \cdot p_2 \right)}{u} \right]
 A_4^{\mathrm{YM}}\left(w,p,\eps\right),
 \nonumber \\
\eq
where the usual definitions of the Mandelstam variables $s=(p_1+p_2)^2$, $t=(p_2+p_3)^2$, $u=(p_1+p_3)^2$ have been
used.

% ----------------------------------------------------------------------------------
\section{Conclusions}
\label{sect:conclusions}

In this paper we have shown that the relation, which relates single trace Einstein-Yang-Mills amplitudes
with one graviton and $(n-1)$ gauge bosons to a linear combination of pure Yang-Mills amplitudes with
$n$ gauge bosons, can be derived from the CHY representation.
The proof relies on the general properties of the building blocks of the 
CHY representations, in particular those of the Parke-Taylor factors. 
In addition, we derived the generalisation for an arbitrary number of gravitons.

\subsection*{Acknowledgements}

L.d.l.C. is grateful for financial support from CONACYT and the DAAD,
A.K. is grateful for financial support from the research training group GRK 1581.

% ----------------------------------------------------------------------------------
% references
\bibliography{/home/stefanw/notes/biblio}

\begin{thebibliography}{10}

\bibitem{Kawai:1985xq}
H.~Kawai, D.~Lewellen, and S.~Tye,
\newblock Nucl.Phys. {\bf B269}, 1 (1986).
%%CITATION = NUPHA,B269,1;%%

\bibitem{Bern:2010ue}
Z.~Bern, J.~J.~M. Carrasco, and H.~Johansson,
\newblock Phys.Rev.Lett. {\bf 105}, 061602 (2010), arXiv:1004.0476.
%%CITATION = ARXIV:1004.0476;%%

\bibitem{Cachazo:2013iea}
F.~Cachazo, S.~He, and E.~Y. Yuan,
\newblock JHEP {\bf 1407}, 033 (2014), arXiv:1309.0885.
%%CITATION = ARXIV:1309.0885;%%

\bibitem{Chen:2010ct}
Y.-X. Chen, Y.-J. Du, and B.~Feng,
\newblock JHEP {\bf 01}, 081 (2011), arXiv:1011.1953.
%%CITATION = ARXIV:1011.1953;%%

\bibitem{Cachazo:2014nsa}
F.~Cachazo, S.~He, and E.~Y. Yuan,
\newblock JHEP {\bf 01}, 121 (2015), arXiv:1409.8256.
%%CITATION = ARXIV:1409.8256;%%

\bibitem{Cachazo:2014xea}
F.~Cachazo, S.~He, and E.~Y. Yuan,
\newblock JHEP {\bf 07}, 149 (2015), arXiv:1412.3479.
%%CITATION = ARXIV:1412.3479;%%

\bibitem{Stieberger:2014cea}
S.~Stieberger and T.~R. Taylor,
\newblock Phys. Lett. {\bf B739}, 457 (2014), arXiv:1409.4771.
%%CITATION = ARXIV:1409.4771;%%

\bibitem{Stieberger:2015qja}
S.~Stieberger and T.~R. Taylor,
\newblock Phys. Lett. {\bf B744}, 160 (2015), arXiv:1502.00655.
%%CITATION = ARXIV:1502.00655;%%

\bibitem{Stieberger:2015kia}
S.~Stieberger and T.~R. Taylor,
\newblock Phys. Lett. {\bf B750}, 587 (2015), arXiv:1508.01116.
%%CITATION = ARXIV:1508.01116;%%

\bibitem{Stieberger:2015vya}
S.~Stieberger and T.~R. Taylor,
\newblock Nucl. Phys. {\bf B903}, 104 (2016), arXiv:1510.01774.
%%CITATION = ARXIV:1510.01774;%%

\bibitem{Stieberger:2016lng}
S.~Stieberger and T.~R. Taylor,
\newblock (2016), arXiv:1606.09616.
%%CITATION = ARXIV:1606.09616;%%

\bibitem{Chiodaroli:2014xia}
M.~Chiodaroli, M.~Günaydin, H.~Johansson, and R.~Roiban,
\newblock JHEP {\bf 01}, 081 (2015), arXiv:1408.0764.
%%CITATION = ARXIV:1408.0764;%%

\bibitem{Chiodaroli:2015rdg}
M.~Chiodaroli, M.~Gunaydin, H.~Johansson, and R.~Roiban,
\newblock (2015), arXiv:1511.01740.
%%CITATION = ARXIV:1511.01740;%%

\bibitem{Chiodaroli:2016jqw}
M.~Chiodaroli,
\newblock (2016), arXiv:1607.04129.
%%CITATION = ARXIV:1607.04129;%%

\bibitem{Nandan:2016pya}
D.~Nandan, J.~Plefka, O.~Schlotterer, and C.~Wen,
\newblock (2016), arXiv:1607.05701.
%%CITATION = ARXIV:1607.05701;%%

\bibitem{Cachazo:2013gna}
F.~Cachazo, S.~He, and E.~Y. Yuan,
\newblock Phys.Rev. {\bf D90}, 065001 (2014), arXiv:1306.6575.
%%CITATION = ARXIV:1306.6575;%%

\bibitem{Cachazo:2013hca}
F.~Cachazo, S.~He, and E.~Y. Yuan,
\newblock Phys.Rev.Lett. {\bf 113}, 171601 (2014), arXiv:1307.2199.
%%CITATION = ARXIV:1307.2199;%%

\bibitem{Dolan:2013isa}
L.~Dolan and P.~Goddard,
\newblock JHEP {\bf 1405}, 010 (2014), arXiv:1311.5200.
%%CITATION = ARXIV:1311.5200;%%

\bibitem{Dolan:2014ega}
L.~Dolan and P.~Goddard,
\newblock JHEP {\bf 1407}, 029 (2014), arXiv:1402.7374.
%%CITATION = ARXIV:1402.7374;%%

\bibitem{He:2014wua}
Y.-H. He, C.~Matti, and C.~Sun,
\newblock JHEP {\bf 1410}, 135 (2014), arXiv:1403.6833.
%%CITATION = ARXIV:1403.6833;%%

\bibitem{Naculich:2014naa}
S.~G. Naculich,
\newblock JHEP {\bf 1409}, 029 (2014), arXiv:1407.7836.
%%CITATION = ARXIV:1407.7836;%%

\bibitem{Naculich:2015zha}
S.~G. Naculich,
\newblock JHEP {\bf 05}, 050 (2015), arXiv:1501.03500.
%%CITATION = ARXIV:1501.03500;%%

\bibitem{Naculich:2015coa}
S.~G. Naculich,
\newblock JHEP {\bf 09}, 122 (2015), arXiv:1506.06134.
%%CITATION = ARXIV:1506.06134;%%

\bibitem{Weinzierl:2014ava}
S.~Weinzierl,
\newblock JHEP {\bf 1503}, 141 (2015), arXiv:1412.5993.
%%CITATION = ARXIV:1412.5993;%%

\bibitem{Kalousios:2013eca}
C.~Kalousios,
\newblock J.Phys. {\bf A47}, 215402 (2014), arXiv:1312.7743.
%%CITATION = ARXIV:1312.7743;%%

\bibitem{Weinzierl:2014vwa}
S.~Weinzierl,
\newblock JHEP {\bf 1404}, 092 (2014), arXiv:1402.2516.
%%CITATION = ARXIV:1402.2516;%%

\bibitem{Lam:2014tga}
C.~S. Lam,
\newblock Phys. Rev. {\bf D91}, 045019 (2015), arXiv:1410.8184.
%%CITATION = ARXIV:1410.8184;%%

\bibitem{Monteiro:2013rya}
R.~Monteiro and D.~O'Connell,
\newblock JHEP {\bf 1403}, 110 (2014), arXiv:1311.1151.
%%CITATION = ARXIV:1311.1151;%%

\bibitem{Cachazo:2015nwa}
F.~Cachazo and H.~Gomez,
\newblock JHEP {\bf 04}, 108 (2016), arXiv:1505.03571.
%%CITATION = ARXIV:1505.03571;%%

\bibitem{Baadsgaard:2015voa}
C.~Baadsgaard, N.~E.~J. Bjerrum-Bohr, J.~L. Bourjaily, and P.~H. Damgaard,
\newblock JHEP {\bf 09}, 129 (2015), arXiv:1506.06137.
%%CITATION = ARXIV:1506.06137;%%

\bibitem{Baadsgaard:2015hia}
C.~Baadsgaard, N.~E.~J. Bjerrum-Bohr, J.~L. Bourjaily, P.~H. Damgaard, and
  B.~Feng,
\newblock JHEP {\bf 11}, 080 (2015), arXiv:1508.03627.
%%CITATION = ARXIV:1508.03627;%%

\bibitem{Baadsgaard:2015ifa}
C.~Baadsgaard, N.~E.~J. Bjerrum-Bohr, J.~L. Bourjaily, and P.~H. Damgaard,
\newblock JHEP {\bf 09}, 136 (2015), arXiv:1507.00997.
%%CITATION = ARXIV:1507.00997;%%

\bibitem{delaCruz:2015dpa}
L.~de~la Cruz, A.~Kniss, and S.~Weinzierl,
\newblock JHEP {\bf 09}, 197 (2015), arXiv:1508.01432.
%%CITATION = ARXIV:1508.01432;%%

\bibitem{delaCruz:2015raa}
L.~de~la Cruz, A.~Kniss, and S.~Weinzierl,
\newblock JHEP {\bf 11}, 217 (2015), arXiv:1508.06557.
%%CITATION = ARXIV:1508.06557;%%

\bibitem{Sogaard:2015dba}
M.~Søgaard and Y.~Zhang,
\newblock Phys. Rev. {\bf D93}, 105009 (2016), arXiv:1509.08897.
%%CITATION = ARXIV:1509.08897;%%

\bibitem{Bosma:2016ttj}
J.~Bosma, M.~Sogaard, and Y.~Zhang,
\newblock (2016), arXiv:1605.08431.
%%CITATION = ARXIV:1605.08431;%%

\bibitem{Huang:2015yka}
R.~Huang, J.~Rao, B.~Feng, and Y.-H. He,
\newblock JHEP {\bf 12}, 056 (2015), arXiv:1509.04483.
%%CITATION = ARXIV:1509.04483;%%

\bibitem{Cardona:2016gon}
C.~Cardona, B.~Feng, H.~Gomez, and R.~Huang,
\newblock (2016), arXiv:1606.00670.
%%CITATION = ARXIV:1606.00670;%%

\bibitem{Gomez:2016bmv}
H.~Gomez,
\newblock JHEP {\bf 06}, 101 (2016), arXiv:1604.05373.
%%CITATION = ARXIV:1604.05373;%%

\bibitem{Cardona:2016bpi}
C.~Cardona and H.~Gomez,
\newblock JHEP {\bf 06}, 094 (2016), arXiv:1605.01446.
%%CITATION = ARXIV:1605.01446;%%

\bibitem{Cardona:2016wcr}
C.~Cardona and H.~Gomez,
\newblock (2016), arXiv:1607.01871.
%%CITATION = ARXIV:1607.01871;%%

\bibitem{Bjerrum-Bohr:2014qwa}
N.~E.~J. Bjerrum-Bohr, P.~H. Damgaard, P.~Tourkine, and P.~Vanhove,
\newblock Phys.Rev. {\bf D90}, 106002 (2014), arXiv:1403.4553.
%%CITATION = ARXIV:1403.4553;%%

\bibitem{Mason:2013sva}
L.~Mason and D.~Skinner,
\newblock JHEP {\bf 1407}, 048 (2014), arXiv:1311.2564.
%%CITATION = ARXIV:1311.2564;%%

\bibitem{Berkovits:2013xba}
N.~Berkovits,
\newblock JHEP {\bf 1403}, 017 (2014), arXiv:1311.4156.
%%CITATION = ARXIV:1311.4156;%%

\bibitem{Gomez:2013wza}
H.~Gomez and E.~Y. Yuan,
\newblock JHEP {\bf 1404}, 046 (2014), arXiv:1312.5485.
%%CITATION = ARXIV:1312.5485;%%

\bibitem{Adamo:2013tsa}
T.~Adamo, E.~Casali, and D.~Skinner,
\newblock JHEP {\bf 1404}, 104 (2014), arXiv:1312.3828.
%%CITATION = ARXIV:1312.3828;%%

\bibitem{Geyer:2014fka}
Y.~Geyer, A.~E. Lipstein, and L.~J. Mason,
\newblock Phys.Rev.Lett. {\bf 113}, 081602 (2014), arXiv:1404.6219.
%%CITATION = ARXIV:1404.6219;%%

\bibitem{Casali:2014hfa}
E.~Casali and P.~Tourkine,
\newblock JHEP {\bf 04}, 013 (2015), arXiv:1412.3787.
%%CITATION = ARXIV:1412.3787;%%

\bibitem{Geyer:2015bja}
Y.~Geyer, L.~Mason, R.~Monteiro, and P.~Tourkine,
\newblock Phys. Rev. Lett. {\bf 115}, 121603 (2015), arXiv:1507.00321.
%%CITATION = ARXIV:1507.00321;%%

\bibitem{Schwab:2014xua}
B.~U.~W. Schwab and A.~Volovich,
\newblock Phys.Rev.Lett. {\bf 113}, 101601 (2014), arXiv:1404.7749.
%%CITATION = ARXIV:1404.7749;%%

\bibitem{Afkhami-Jeddi:2014fia}
N.~Afkhami-Jeddi,
\newblock (2014), arXiv:1405.3533.
%%CITATION = ARXIV:1405.3533;%%

\bibitem{Zlotnikov:2014sva}
M.~Zlotnikov,
\newblock JHEP {\bf 1410}, 148 (2014), arXiv:1407.5936.
%%CITATION = ARXIV:1407.5936;%%

\bibitem{Kalousios:2014uva}
C.~Kalousios and F.~Rojas,
\newblock JHEP {\bf 01}, 107 (2015), arXiv:1407.5982.
%%CITATION = ARXIV:1407.5982;%%

\bibitem{White:2014qia}
C.~White,
\newblock Phys.Lett. {\bf B737}, 216 (2014), arXiv:1406.7184.
%%CITATION = ARXIV:1406.7184;%%

\bibitem{Monteiro:2014cda}
R.~Monteiro, D.~O'Connell, and C.~D. White,
\newblock JHEP {\bf 12}, 056 (2014), arXiv:1410.0239.
%%CITATION = ARXIV:1410.0239;%%

\bibitem{delaCruz:2016wbr}
L.~de~la Cruz, A.~Kniss, and S.~Weinzierl,
\newblock Phys. Rev. Lett. {\bf 116}, 201601 (2016), arXiv:1601.04523.
%%CITATION = ARXIV:1601.04523;%%

\bibitem{White:2016jzc}
C.~D. White,
\newblock (2016), arXiv:1606.04724.
%%CITATION = ARXIV:1606.04724;%%

\bibitem{Kleiss:1988ne}
R.~Kleiss and H.~Kuijf,
\newblock Nucl. Phys. {\bf B312}, 616 (1989).
%%CITATION = NUPHA,B312,616;%%

\bibitem{Bern:2008qj}
Z.~Bern, J.~J.~M. Carrasco, and H.~Johansson,
\newblock Phys. Rev. {\bf D78}, 085011 (2008), arXiv:0805.3993.
%%CITATION = 0805.3993;%%

\bibitem{BjerrumBohr:2010ta}
N.~Bjerrum-Bohr, P.~H. Damgaard, B.~Feng, and T.~Sondergaard,
\newblock Phys.Rev. {\bf D82}, 107702 (2010), arXiv:1005.4367.
%%CITATION = ARXIV:1005.4367;%%

\bibitem{BjerrumBohr:2010hn}
N.~Bjerrum-Bohr, P.~H. Damgaard, T.~Sondergaard, and P.~Vanhove,
\newblock JHEP {\bf 1101}, 001 (2011), arXiv:1010.3933.
%%CITATION = ARXIV:1010.3933;%%

\end{thebibliography}
\bibliographystyle{/home/stefanw/latex-style/h-physrev5}

\end{document}